\documentclass[preprint,amsmath,amssymb,aps,showkeys,showpacs]{revtex4}
\usepackage[colorlinks=true, pdfstartview=FitV, linkcolor=blue, citecolor=red, urlcolor=magenta]{hyperref}
\usepackage{graphicx}
\usepackage{latexsym}
\usepackage{amsmath}
\usepackage{amsfonts}
\usepackage{amssymb}
\usepackage{verbatim}
\usepackage{dcolumn}
\usepackage{amssymb}
\usepackage{bm}
\usepackage{float}
\usepackage{subfigure}
\usepackage[english]{babel}
\newcommand{\be}{\begin{equation}}
\newcommand{\ee}{\end{equation}}
\newcommand{\bea}{\begin{eqnarray}}
\newcommand{\eea}{\end{eqnarray}}

\begin{document}

\newcommand{\NITK}{
\affiliation{Department of Physics,\\ National Institute of Technology Karnataka, Surathkal  575 025, India}
}

\title{Effect of Dark Energy in Geometrothermodynamics and Phase Transitions of Regular Bardeen AdS Black Hole}
\author{C.L. Ahmed Rizwan}
\email{ahmedrizwancl@gmail.com}
\NITK
\author{A. Naveena Kumara}
\email{naviphysics@gmail.com}
\NITK
\author{K.V. Rajani}
\email{rajanikv10@gmail.com}
\NITK
\author{Deepak Vaid}
\email{dvaid79@gmail.com}
\NITK
\author{K.M. Ajith}
\email{ajith@nitk.ac.in}
\NITK

\begin{abstract}
We investigate thermodynamics and geometrothermodynamics of regular Bardeen-AdS black holes with quintessence. The thermodynamics of the black hole is studied using temperature-entropy ($T-S$) , Pressure-Volume ($P-v$), and Gibbs energy plots, which indicate a critical behaviour. This is also confirmed from the divergence of specific heat against entropy, which shows a second-order phase transition. Moreover,  we observe that the quintessence state parameter $\omega$ shifts the transition point to lower entropy values.  Using the concept of thermodynamic Ruppeiner and Weinhold geometry, we calculated the thermodynamic curvature scalar $R_R$ and $R_W$ in the quintessence dark energy regime ($\omega =-2/3$). While these curvature scalars enable us to identify the critical behavior, they do not show divergence at the phase transition points observed in specific heat studies. To resolve this puzzle, we have adopted the method of geometrothermodynamics proposed by Hernando Quevedo. Choosing a Legendre invariant ‘Quevedo’ metric, the curvature scalar $R_Q$ shows singularity at the same point as seen in the specific heat divergence.

\end{abstract}

\maketitle


\section{Introduction}
\label{sec:intro}

In recent years, black hole thermodynamics has become an active area of research in theoretical physics. Among several motivations, the main attraction lies in the fact that the black hole is the best system to seek the aspects of quantum gravity, and the thermodynamic study will reveal its microscopic structure. Introduction of correspondence between classical gravitational theory in AdS space and strongly coupled conformal field theory on its boundary by Maldacena in his seminal paper \citep{Maldacena1999} made the thermodynamic study of asymptotically AdS black holes more interesting. The black hole thermodynamics in anti-de Sitter (AdS) space is different from asymptotically Minkowskian spacetime. The AdS space acts like a thermal cavity, and a black hole can exist in a stable equilibrium with radiation. But there is a minimum Hawking temperature (critical temperature) below which only thermal radiation exists. Above this temperature, two types of black hole solutions exist, a smaller black hole with negative specific heat capacity and a larger black hole with positive specific heat capacity. At the critical temperature, Hawking-Page phase transition takes place between thermal radiation and large black hole  \citep{Hawking1983}. In the AdS-CFT perspective, Hawking-Page phase transition is understood as confinement/deconfinement phase transition in gauge theory \citep{Witten:1998zw}.

Realizing the importance of thermodynamics in AdS space, the Reissner-Nordstr{\"o}m and the Kerr-Newman black holes in the AdS background were studied. The small-large black hole phase transition found in RN-AdS had a close resemblance to van der Waals liquid-gas system \citep{Chamblin1999, Chamblinb1999, Caldarelli2000}. More clarity on this isomorphism was obtained by identifying the cosmological constant as thermodynamic pressure and by expanding the first law by including a PdV term \citep{Dolan2011a, Dolan2011b, Kubiznak:2012wp, Kastor2009}. Recently, thermodynamics of various black holes in this extended phase space was studied, and the similarity with van der Waals liquid-gas system was found to be universal  \citep{Gunasekaran2012, BelhajChabab2012, SPALLUCCI2013, Altamiranokubi2013, Zhao2013, Hendi2013, SChen2013}.

After Albert Einstein's theory of gravity based on differential geometry became a great success, the method of differential geometry was identified as a mathematical language for various gauge fields. It was Gibbs \citep{gibbs1948collected} in the later part of the 19th century and Caratheodory \citep{Caratheodory1909} in 1909, to use these ideas of differential geometry in classical thermodynamics. Hermann \citep{hermann1973geometry} and Mrugala \citep{MRUGALA1978419} applied differential geometry to the thermodynamic phase space making use of its contact structure. Then Weinhold \citep{weinhold76} and later Ruppeiner \citep{Ruppeiner79,Ruppeiner95} constructed thermodynamic metric to study phase transitions and microscopic interactions in thermodynamic systems. Geometrothermodynamics is another geometric formalism for the classical thermodynamics developed by H.Quevedo \citep{Quevedo2007}. Recently, a brief history of metric geometry of thermodynamics was written by Ruppeiner \citep{Ruppeiner2016}. From these geometric formalisms, a metric is defined on equilibrium thermodynamic state space, and the thermodynamic curvature scalar encodes the information about the microscopic interactions. The curvature scalar is proportional to the correlation volume, and its sign tells the nature of microscopic interactions being attractive or repulsive \citep{Ruppeiner2010}. The phase transition of the system can be seen in the divergence behavior of this curvature scalar near the critical point. Thermodynamic geometry  is applied to van der Waals gas and  different statistical models including magnetic models \citep{MagMrugala1989,vdwJanyszek90,Brody95,FerroDolan98,nonDolan2002,IsingJanke2002}.
 
Considering black hole as a thermodynamic system, the geometric formalism is used to study the critical behavior of black holes during phase transitions \citep{Ruppeinerb2008,Aman2003,bTzSarkar2006,SHEN2007,TSarkar2008,Sahaycritical2010,Banerjee2010,BANERJEE2011,Akbar2011, Niu2012,BellucciEYM2012,Lala2012,Wei2013,Yi5d2013,
Suresh2014,Zhang2015,CVRMansoori2015,
ExtndZhangRN2015,SheykhiEMd2015,SoroushfR2016,
lifR2016,Sahay2017,Chaturvedi2017,HOSSEINIMANSOORI2016298}. But there were inconsistencies in the position of the critical point, as specific heat diverges at a point different from where scalar curvature diverges \citep{GTDJanke2010, MismatchWei2012, Suresh2014}. The Legendre invariance was found to be the key factor behind these discrepancies. Taking Legendre invariance into account, a metric was constructed by Quevedo et al. \citep{Quevedo2007, Quevedo12008, Quevedo2008}, which resolved the issue. Quevedo's formalism named as Geometrothermodynamics(GTD) is applied to various black holes \citep{GTDMyers2013,GTDquevado2012,Tharanath2015,GTDSanchez2016,GTDEMDQuevado16, GTDHu2017,GTDScalrtensor2018} including regular black holes. A black hole without singularity at the origin possessing an event horizon is called a regular or non-singular black hole. It was Bardeen \citep{bardeen1968non} in 1968 who constructed a black hole solution with regular non-singular geometry with an event horizon for the first time. Later, several such regular black hole solutions were constructed \citep{Hayward, AyonBeato:1998ub}. The thermodynamic properties of  Bardeen regular black holes were studied in \citep{MAkbar2012}.
  
The accelerated expansion of the universe is due to the presence of an exotic field called Dark energy. Quintessence is one among the different models for dark energy  \citep{Ford1987, Kiselev2003, Shinji2013}. The cosmic source for inflation has the equation of state $p_q=\omega \rho _q$ ($-1< \omega < -1/3$), and $\omega =-2/3$ corresponds to quintessence dark energy regime. The energy density for quintessence has the form $\rho _q=-\frac{a}{2}\frac{3\omega}{r^{3(\omega+1)}}$, which is positive for usual quintessence. Several attempts have been made to explore the effects of quintessence on the black hole, with Kiselev's \citep{Kiselev2003} phenomenological approach being the notable one. Phase transitions in black holes surrounded by quintessence are widely studied. Thermodynamics of Reissner-Nordstr{\"o}m and regular  black holes surrounded by quintessence were investigated in \citep{WeiYi2011,QuienRNThomas2012,THARANATH2013,LI2014,Fan2017,Saleh2018,Rodrigue2018lzp}.
Thermodynamic geometry and geometrothermodynamics for different regular black holes were studied in \citep{Tharanath2015}. Thermodynamic geometry of charged AdS black hole surrounded by quintessence can be found in \citep{ShaoWenWei2018}.
 
This paper is organized as follows. In section  \ref{sec:thermo}, thermodynamics of regular Bardeen black hole surrounded by quintessence is studied in the extended phase space. In the next section (\ref{sec:tg}), thermodynamic geometry for the black hole is constructed using Weinhold and Ruppeiner metric, followed by geometrothermodynamics. Conclusion is written in the final section (\ref{sec:conclu}).


\section{Thermodynamics of Regular-Bardeen black hole}
\label{sec:thermo}
The action for regular black hole solution  coupled with  nonlinear electrodynamics in AdS space can be expressed as  \citep{AyonBeato:1998ub}
\begin{align}
S=\int d^4 x \sqrt{-g}\left(\frac{1}{16 \pi}R+\frac{1}{8\pi}\frac{3}{l^2} -\frac{1}{4 \pi}\mathcal{L}\left(\mathcal{F}\right) \right).
\end{align}
In which $R$ is the Einstein scalar, $l$ the AdS radius and $ \mathcal{L}\left(\mathcal{F}\right)$ is the Lagrangian for a nonlinear electrodynamics source,
\begin{align}
\mathcal{L}\left(\mathcal{F} \right)=\frac{3M}{\beta^3}\left(\frac{ \sqrt{4\beta ^2 \mathcal{F}}}{1+\sqrt{4\beta ^2 \mathcal{F}}}\right).
\end{align}
Here $\mathcal{F}=F_{\mu\nu}F^{\mu\nu}$, ${F}_{\mu\nu}$ is the electromagnetic field strength, and $\beta$ is the magnetic monopole charge associated with it.
The spherically symmetric solution for the action is obtained as \citep{AyonBeato:1998ub}
\begin{align*}
ds^2&= -\left(1-\frac{2 \mathcal{M}(r)}{r} +\frac{r^2}{l^2}\right)dt^2+ \frac{dr^2}{\left(1-\frac{2 \mathcal{M}(r)}{r}+\frac{r^2}{l^2} \right)} +r^2d\theta^2 +r^2 \sin^2 \theta d\phi^2,
\end{align*}
where $\mathcal{M}(r)= \frac{Mr^3}{(r^2+\beta ^2)^{3/2}}$.
\paragraph{}Presence of quintessence throughout the universe makes  it important to probe its effects on the black holes. By using Kislev's phenomenological model, one can construct regular-Bardeen  black hole surrounded by quintessence. According to this model,
the quintessence comes from a fluid with the energy-momentum tensor
\begin{align*}
T^r_r=T^t _t&=\rho_q,\\
T^{\theta}_{\theta}=T^{\phi}_{\phi}&=-\frac{1}{2}\rho_q\left(3 \omega+1\right),\\
\textrm{and}\quad\rho _q&=-\frac{a}{2}\frac{3\omega}{r^{3(\omega+1)}},
\end{align*}
$\omega$ and $a$ are the state parameter and the normalization constant related to quintessence energy density $\rho_q$.
 Solving the Einstein equations, we obtain the metric for a regular-Bardeen AdS black hole with quintessence \citep{Saleh2018,Fan2017,LI2014,Kiselev2003},
 \begin{align*}
ds^2&= -f(r)dt^2+ \frac{dr^2}{f(r)} +r^2d\theta^2 +r^2 \sin^2 \theta d\phi^2,
\end{align*}
where
 \begin{align*}
 f\left(r\right)= \left(1-\frac{2 \mathcal{M}(r)}{r} +\frac{r^2}{l^2}-\frac{a}{r^{3\omega +1}}\right).
 \end{align*}
In the extended phase space cosmological constant is considered as thermodynamic pressure.  
\begin{eqnarray}
P=-\frac{\Lambda}{8\pi}, ~~\quad \Lambda=-\frac{3}{l^2}. 
\end{eqnarray}
On the event horizon $r_h$, $f(r_h)=0$ which gives the mass M corresponding to the above metric, 
\begin{eqnarray}
M =\frac{1}{6} r_h^{-3 (1 + \omega)} (\beta ^2 + r_h^2)^{3/2} [-3 a + r_h^{1 + 3 \omega} (3 + 8 P \pi r_h^2)].
\end{eqnarray}

We can express the mass of the black hole as a function of entropy using the area law $S=\pi r_h ^2$ as follows,

\begin{align}
M=&\frac{1}{6 \sqrt{\pi }}\left[ \left(\pi  \beta^2+S\right)^{3/2} S^{-\frac{3}{2}  (\omega +1)} \left((8 P S+3) S^{\frac{3 \omega }{2}+\frac{1}{2}}-3 a \pi ^{\frac{3 \omega }{2}+\frac{1}{2}}\right)\right] .
\label{mass}
\end{align}
First law of black hole thermodynamics must be modified to include quintessence as follows,
\begin{eqnarray}
dM=TdS+\Psi d\beta+VdP+\mathcal{A}d a .\label{eqn:First law}
\end{eqnarray}

Where $\Psi$ is the potential conjugate to the magnetic charge $\beta$ and $\mathcal{A}$ is a quantity conjugate to quintessence parameter $a$.
\begin{equation}
\mathcal{A}=\left( \frac{\partial M}{\partial a}\right) _{S,\beta , P}=-\frac{1}{2r_h^{3\omega}}.
\end{equation}
The  quintessence parameters $\mathcal{A}$ and $a$ are introduced similar to pressure-volume term to make the first law consistent with the Smarr relation,
\begin{equation}
M = 2 T S + \Psi \beta -2P V+ (3\omega + 1)\mathcal{A} a.
\end{equation}
This can be obtained through Euler's theorem as the mass of the black hole equation (\ref{mass}) is a homogenous function.
The quintessence parameter $a$ being a thermodynamic variable contributes to the internal energy of black holes and hence to the thermodynamics. The appearance of $a$ and its conjugate variable $\mathcal{A}$ in first law and Smarr relation justifies our motivation to study its effect in a phase transition. It is also worth recalling that these modifications due to quintessence is parallel to the extension by identifying cosmological constant as thermodynamic pressure.  These two identifications plays a crucial role in the critical phenomena of AdS black holes.

We can derive Hawking temperature from the first law (equation \ref{eqn:Hawking temperature}), which can be combined with the area law to  obtain equation of state (equation \ref{eqn of state}).
\begin{align}
T&=\left( \frac{\partial M}{\partial S} \right)_{\Psi,P,a}\\
&=\textstyle{\frac{\left(\beta ^2+\frac{S}{\pi }\right)^{\frac{1}{2}}}{4} S^{-\frac{3 \omega }{2}-\frac{5}{2}} \left[3 a \pi ^{\frac{3 \omega }{2}+\frac{1}{2}} \left(\pi  \beta ^2 (\omega +1)+S
   \omega \right)+S^{\frac{3 \omega }{2}+\frac{1}{2}} \left(-2 \pi  \beta ^2+8 P S^2+S\right)\right]}\label{eqn:Hawking temperature}.\\
P=&\frac{1}{8 \pi }\left[ \frac{8 \pi  T}{\sqrt{4 \beta ^2+v^2}}+32 \beta ^2 v^{-3 \omega -5} \left(v^{3 \omega +1}-3 a 8^{\omega } (\omega +1)\right)-3 a 8^{\omega +1} \omega  v^{-3 (\omega +1)}-\frac{4}{v^2}\right] \label{eqn of state}.
\end{align}
\begin{figure*}[ht!]
	\subfigure[]
	{
		\includegraphics[width=0.45\textwidth]{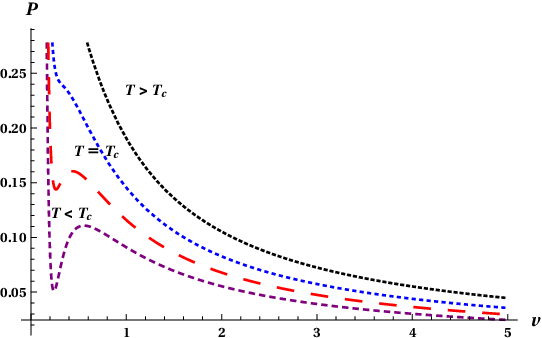}
		\label{PV}
	}
	\subfigure[]
	{
		\includegraphics[width=0.45\textwidth]{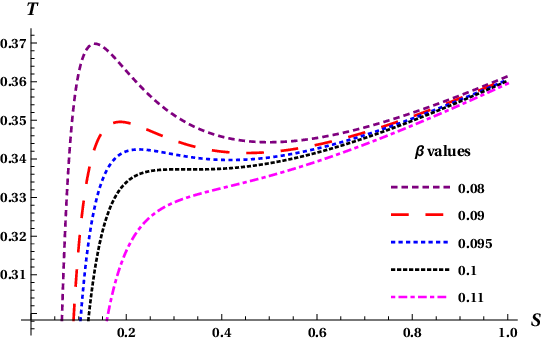}
		\label{TS}
	}
\\
  \caption{To the left we have $P-v$ diagram for regular AdS black hole surrounded by quintessence ($a=0.07$, $\beta=0.1$, $\omega =-2/3$,  $T_c=0.36$). In the right side $T-S$ plot for different values of $\beta$ is shown.}\label{PVTS}
\end{figure*}



\noindent where  $v=2r_h$ is  specific volume. Using the above equations  the $P-v$ and $T-S$ curves are plotted in figure \ref{PV} and \ref{TS}. Below a critical temperature $T_C$, $P-v$ isotherm has three branches corresponding to small, intermediate, and large black holes. This behavior is quite similar to liquid/gas transition van der Waals fluids. Two plots clearly show critical phenomena around the critical points. The critical points are obtained from the conditions, 
\begin{equation}
\frac{\partial P}{\partial v}=0 \quad, \quad \quad \frac{\partial^2 P}{\partial v^2}=0.\label{criticalform}
\end{equation}

In the absence of quintessence, the critical volume $(v_c)$, critical temperature $(T_c)$ and critical pressure $(P_c)$ of regular Bardeen-AdS black hole are obtained, which are as follows,
 \begin{align*}
 v_c&=2\sqrt{2}\beta \sqrt{ 2+\sqrt{10} }\quad,\quad
 T_c=\frac{25 \left(13 \sqrt{10}+31\right)}{432\pi \beta \left(2 \sqrt{10}+5\right)^{3/2} }\quad,\\
 P_c&=\frac{5 \sqrt{10}-13}{432 \pi {\beta}^2}
 \end{align*}
 Using the critical quantities, we can calculate $\frac{P_c v_c}{T_c}$ ratio.
 \begin{align*}
\frac{ P_c v_c}{T_c}&= \frac{   \left(-26+10 \sqrt{10}\right)\left(5+2 \sqrt{10}\right)^{3/2}\sqrt{2 \left(2+\sqrt{10}\right)}}{775+325 \sqrt{10}}
\end{align*} 
which is numerically equal to 0.381931. The ratio $\frac{P_c v_c}{T_c}$ gives the critical compressibility factor, which is the measure of deviation from ideal gas behaviour. This factor tells about the intermolecular forces in gases, and for an ideal gas, it is one. For van der Waals fluid, this value is $3/8$, which is a universal number. Negative deviation of this factor from unity in real gas indicates an attractive microstructure interaction, unlike in ideal gas where there exists no interaction.  In the case of Reissner-Nordstr{\"o}m  AdS black hole,  this ratio matches with that of a  Van der Waals gas. This gives another signature of van der Waals-like interactions in the microstructure and critical phenomena in charged AdS black holes. The inclusion of the quintessence parameter increases the ratio, which implies a change in the microscopic interaction and in phase structure.

\begin{figure*}[ht!]
	\subfigure[]
	{
		\includegraphics[width=0.45\textwidth]{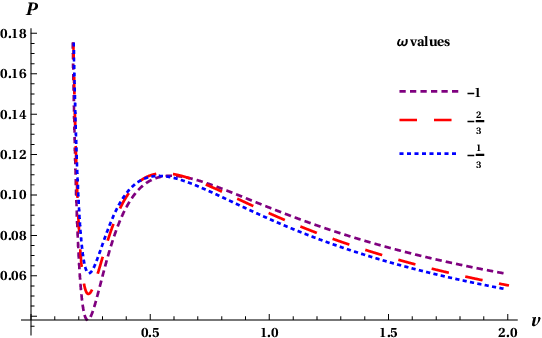}
		\label{omegaonPV}
	}
	\subfigure[]
	{
		\includegraphics[width=0.45\textwidth]{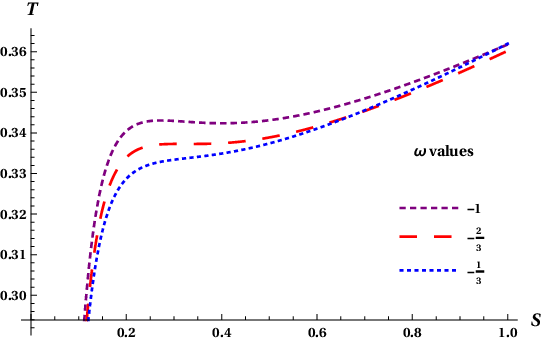}
		\label{omegaonTS}
	}
\\
  \caption{Effect of quintessence parameter $\omega$ on $P-v$ isotherms (figure \ref{omegaonPV}) and temperature (figure \ref{omegaonTS}). }\label{omegaonPVTS}
\end{figure*}
Through the critical values of temperature, pressure and volume, we can find how presence of quintessence affects the phase transition. As the analytic expression is difficult to obtain, the critical quantities are obtained numerically for the state parameter  $\omega=-1,-2/3,-1/3$ (table  \ref{tab:table1}). Effect of quintessence in the $P-v$ isotherms and $T-S$ plots are shown in figure \ref{omegaonPVTS}. Increase in the value of $\omega$ from $-1$ to $0$, leads to decrease in the $\frac{P_c V_c}{T_c}$ ratio, which approaches to $3/8$.

\begin{table}[ht!]
\caption{\label{tab:table1}%
Critical points are found using equation (\ref{criticalform}) with quintessence state parameter $\omega =-1,-2/3,-1/3$. The ratio $\frac{P_c v_c}{T_c}$ is calulated for each case.
}
\begin{center}

\begin{tabular}{|l|l|l|l|l|}

\hline
\textrm{$\omega$}&
\textrm{$P_c$}&
\textrm{$v_c$}&
\textrm{$T_c$}&
\textrm{$\frac{P_c v_c}{T_c}$}\\
\hline\hline
-1 & 0.2155 & 0.6426 & 0.3485 & 0.3973\\
$-2/3$ & 0.2073 & 0.6422 & 0.3376 & 0.3945\\
$-1/3$& 0.1926 & 0.6426 & 0.3241 & 0.3819\\
\hline\noalign{\smallskip}
\end{tabular}
\end{center}
\end{table}

In statistical mechanics, a phase transition is characterized by divergences in second moments like specific heat, compressibility, and susceptibility.
Hence to study more details of phase transition we focus on heat capacity of the system. Sign of heat capacity tells about the thermodynamic stability of black hole, which is positive for stable and negative for unstable. The heat capacity at 
constant pressure is given by,
\begin{align*}
C_P=&T\left( \frac{\partial S}{\partial T}\right)_P\\
=&\frac{2 S \left(\pi  \beta ^2+S\right) \left(S \left(\sqrt{\pi } (8 P S+1)-2 a \sqrt{S}\right)+\beta ^2 \left(\pi  a \sqrt{S}-2 \pi
   ^{3/2}\right)\right)}{\sqrt{\pi } \left(\beta ^4 \left(8 \pi ^2-3 \pi ^{3/2} a \sqrt{S}\right)+S^2 (8 P S-1)+4 \pi  \beta ^2 S\right)}.
\end{align*}
\begin{figure*}[ht!]
\subfigure[]
{
\includegraphics[width=0.33\textwidth]{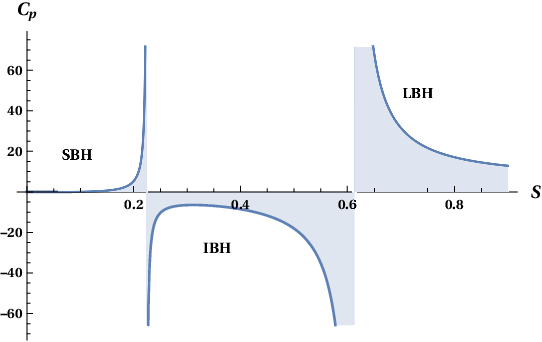}
\label{sub1}
}
\subfigure[]
{
\includegraphics[width=0.3\textwidth]{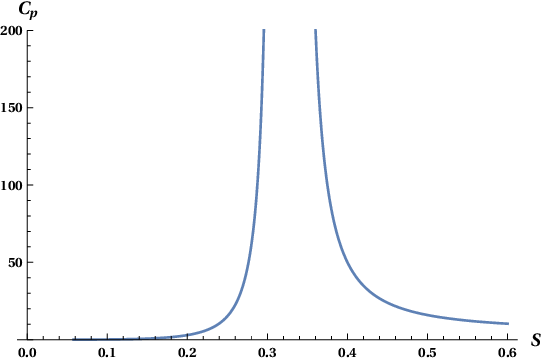}
\label{sub2}
}
\subfigure[]
{
\includegraphics[width=0.3\textwidth]{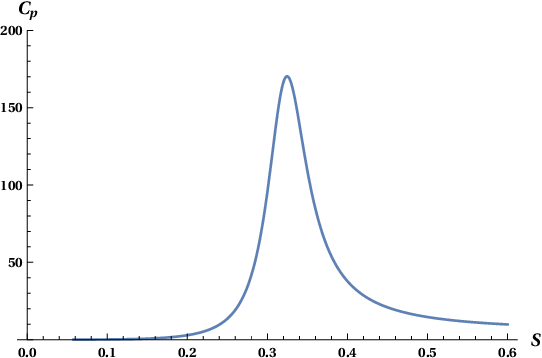}
\label{sub3}
}
\\
\caption{ Specific heat versus entropy diagram for regular AdS black hole surrounded by quintessence ($a=0.07$, $\beta=0.1$ , $\omega=-\frac{2}{3}$). (\ref{sub1}) for $P=Pc $, (\ref{sub2})  for $P<Pc$, (\ref{sub3}) for $P>Pc$.}\label{CS}
\end{figure*} 

\begin{figure}[ht!]
\subfigure[]
{
\includegraphics[width=0.47\textwidth]{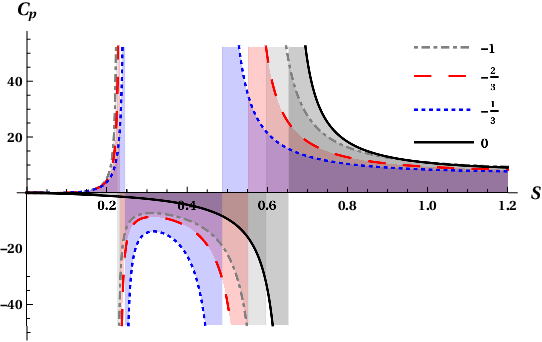}
		\label{CSomegas}
}
\subfigure[]
{
\includegraphics[width=0.47\textwidth]{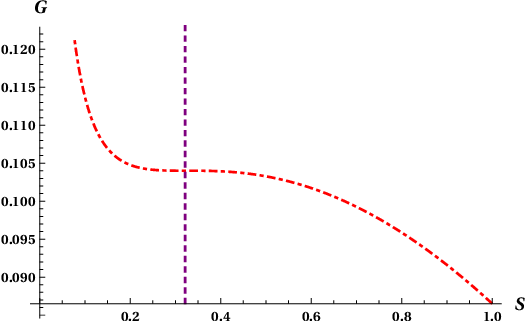}
		\label{Gibbs}
		}
		\\ 
\caption{Specific heat at constant pressure $C_P$ versus entropy $S$ is plotted for different $\omega$ values in figure \ref{CSomegas}. In figure \ref{Gibbs} Gibbs free energy $G$ vs entropy S is plotted for a fixed value of $\omega =-\frac{2}{3}$. In  both plots $ a=0.07$ and $\beta=0.1$.}
\end{figure}

$C_P-S$ plot is obtained from this equation, which shows critical behavior (figure \ref{CS}) below certain pressure $(P_c)$. These plots show that below the critical pressure $P<P_c$, there are two singular points, which reduce to one when $P=P_c$, and above $P>P_c$, these divergence disappears. In figure \ref{sub2}, there are three distinct regions separated by two singular points. The Small black hole (SBH) and large black hole (LBH) regions with positive specific heat, and the intermediate black hole (IBH) with negative specific heat.  As the positive specific heat regions are thermodynamically stable, a phase transition occurs between a small black hole and a large black hole. This result is analogous to Reissner-Nordstr{\"o}m AdS black holes surrounded by quintessence\citep{QuienRNThomas2012}. In figure \ref{CSomegas}, we observe that the quintessence state parameter $\omega$ shifts the SBH-LBH transition to lower entropy values. The specific heat  plotted with $\omega=-1,-\frac{1}{3},-\frac{2}{3}$ and $0$ shows the deviation.   As the initial and final phases have positive specific heat for $\omega\neq 0$, the system can exhibit a \emph{reentrant phase transition} (RPT). However, when $\omega = 0$, the intermediate region vanishes, and only two regions exist, one with negative and the other with positive specific heat. Therefore we can conclude that it is the  quintessence field surrounding the black hole that leads 
to \emph{reentrant phase transitions} which is absent in regular Bardeen AdS black hole.

For more details in the thermodynamic instability, we study  Gibbs free energy $\left(G=M-T S\right)$. At fixed quintessence parameter $\omega=-2/3$, the Gibbs free energy in terms of entropy S and pressure P is given by,
\begin{align}
G\left(S,P\right)=\frac{\sqrt{b^2+\frac{S}{\pi }} \left(b^2 \left(4 \pi  (4 P S+3)-9 \sqrt{\pi } a \sqrt{S}\right)+s (3-8 P S)\right)}{12 S}.
\end{align}
Like in the case of a specific heat plot, the Gibbs free energy also indicates a phase transition. A swallowtail behavior is expected when G is plotted against temperature T, but inverting equation(\ref{eqn:Hawking temperature}) is difficult.
However, the Gibbs free energy plotted as a function of entropy at critical pressure $P_c=0.207$ shows a local minimum near the critical value of S.


\section{Thermodynamic Geometry}
\label{sec:tg}
In this section, we investigate thermodynamic phase transitions based on geometric formalism proposed by Weinhold, Ruppeiner, and Quevedo. The thermodynamic geometry is a possible tool to explore thermodynamic phase transitions from a microscopic point of view. The thermodynamic scalar curvature \emph{R} is directly proportional to the correlation volume of the system $R\propto  \xi ^d $, where $d$ is spatial dimensionality. The divergent behavior of curvature scalar plotted against entropy reflects the existence of critical points corresponding to a thermodynamic phase transition.
\subsection{Weinhold Geometry}
The Weinhold metric is defined ad-hoc in the thermodynamic equilibrium space as the Hessian of the internal energy M, 
\begin{align*}
ds_W^{2}&=g_{ij}^{W}dx^{i}dx^{j}= \partial _i \partial _j M(S,N^a) dx^{i}dx^{j} ~~,~~(i,j=1,2)
\end{align*}
where  $N^a$ represents any other thermodynamic extensive variables. Here, mass M is the function of entropy $S$ and extensive variable  $\beta$, which is the monopole charge. A  Hessian  is defined as a square matrix containing the second derivative of energy with respect to the entropy and other extensive parameters\citep{weinhold1975metric,weinhold76},
\begin{align*}
g^W=\begin{bmatrix}
     M_{,SS} & M_{,S \beta}\\
     M_{, \beta S} & M_{,\beta \beta}
    \end{bmatrix}.
\end{align*}
Using the expression for mass of the black hole  (equation \ref{mass}), the components of metric tensor turns out to be,

\begin{align}
g_{SS}&=\frac{\beta ^4 \left(8 \pi ^2-3 \pi ^{3/2} a \sqrt{S}\right)+S^2 (8 P S-1)+4 \pi  \beta ^2 S}{8 \sqrt{\pi } S^3 \sqrt{\pi  \beta ^2+S}} \label{Weinhold metric1}\\ 
g_{S\beta}&=g_{\beta S}=\frac{\beta  \left(\beta ^2 \left(3 \sqrt{\pi } a \sqrt{S}-6 \pi \right)+S (8 P S-3)\right)}{4 S^2 \sqrt{\beta ^2+\frac{S}{\pi }}} \label{Weinhold metric2}\\ 
g_{\beta \beta}&=\frac{\left(2 \pi  \beta ^2+S\right) \left(\sqrt{\pi } (8 P S+3)-3 a \sqrt{S}\right)}{2 S \sqrt{\pi  \beta ^2+S}}.
\label{Weinhold metric3}
\end{align}

\begin{figure*}[ht!]
\subfigure[]
{
\includegraphics[width=.3\textwidth]{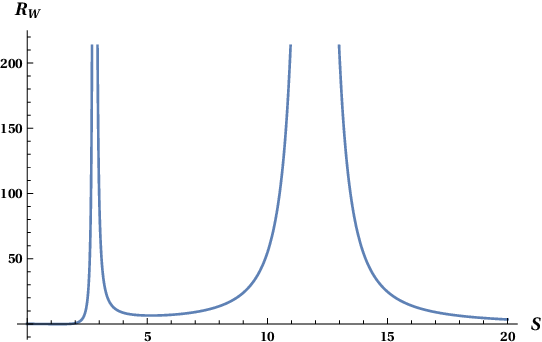}
\label{WRS1}
}
\subfigure[]
{
\includegraphics[width=0.3\textwidth]{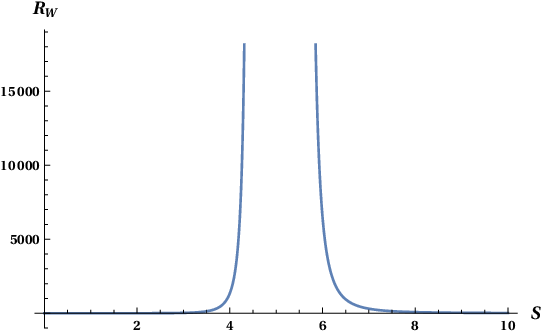}
\label{WRS2}
}
\subfigure[]
{
\includegraphics[width=0.3\textwidth]{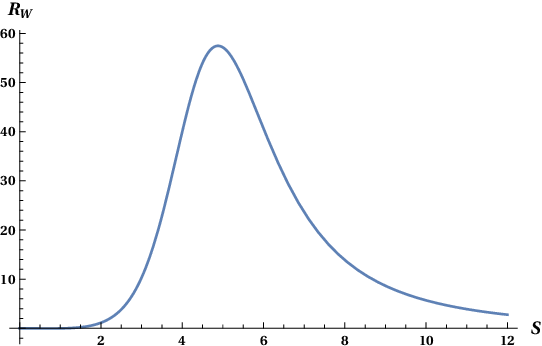}
\label{WRS3}
}
\\
\caption{Curvature divergence plots for Weinhold metric. In all three plots quintessence parameter and monopole charge are fixed ( $a=0.5$ and $\beta =1$). The value of pressure is $P=0.01$ in \ref{WRS1}, $P=0.01264$ in \ref{WRS2} and $P=0.0141$ in \ref{WRS3}.}\label{Weinhold RS}
\end{figure*}

From metric tensor $g^W_{ij}$, one can calculate curvature scalar, which is found to be a complicated expression, $R_W(S,P,b,\omega,a)$. Plotting the curvature $R_W$ versus entropy $S$, we have studied its divergence behavior, which occurs at multiple points (figure \ref{Weinhold RS}).
Even at the critical point ($P_c$=0.207 for $a$=0.07 and $\beta =0.1$),  $R_W$ shows multiple divergences (figure \ref{Criticalw}) which are different from that of the critical value of entropy $(S)$ observed in specific heat plots. From these randomly located diverging points, we can infer only the critical behavior of the system but not the exact phase transition points. As there is no agreement between the divergence points in Weinhold geometry and specific heat study, next, we focus on Ruppeiner geometry.



\subsection{Ruppeiner Geometry}
The Ruppeiner metric is defined as a Hessian of the entropy function $S$ of the system instead of the internal energy $M$ as in the Weinhold case. But one can transform the Ruppeiner metric, which is a function of $M$ and $\beta$ originally,  to the same coordinate system used in Weinhold metric, i.e., $S$ and $\beta$. Technically, their geometries are related to each other conformally  \citep{Ruppeiner79,Ruppeiner95,Ruppeinerb2008,Ruppeiner2010}.

The Ruppeiner metric in the thermodynamic space states is given as ,
\begin{align*}
 g_{ij}^{R}&=-\partial_{i}\partial_{j}S\left(M,x^{\alpha}\right)~~~~~~(i,j=1,2)
\end{align*}
\begin{align*}
g^R=\begin{bmatrix}
                   S_{,M M} & S_{,M \beta}\\
                   S_{,\beta M} & S_{,\beta \beta}
    \end{bmatrix}.
\end{align*}
Because of the conformal property, the line elements in Ruppeiner and Weinhold formalism are related as
\begin{equation}
dS^2 _R=-\frac{dS^2 _W}{T}.
\end{equation}
Using (\ref{Weinhold metric1}), (\ref{Weinhold metric2}), (\ref{Weinhold metric3}) and (\ref{eqn:Hawking temperature}) the components of Ruppeiner metric tensor are easily obtained as , 

\begin{align}
g_{SS} &=\frac{\sqrt{\pi }\left(b^4 \left(8 \pi ^2-3 \pi ^{3/2} a \sqrt{S}\right)+4 \pi  b^2 S+S^2 (8 P S-1)\right)}{2 S \left(\pi  b^2+S\right)\left(b^2 \left(\pi  a \sqrt{S}-2 \pi ^{3/2}\right)+S \left(\sqrt{\pi } (8 P S+1)-2 a \sqrt{S}\right)\right)}\\
g_{S\beta} &=g_{\beta S}=\frac{\pi ^{3/2} b \left(b^2 \left(6 \pi -3 \sqrt{\pi } a \sqrt{S}\right)+S (3-8 P S)\right)}{\left(\pi  b^2+S\right) \left(b^2 \left(2 \pi ^{3/2}-\pi  a \sqrt{S}\right)-S \left(\sqrt{\pi } (8 P S+1)-2 a \sqrt{S}\right)\right)}\\
g_{\beta \beta}&=\frac{2 \pi  S \left(2 \pi  b^2+S\right) \left(\sqrt{\pi } (8 P S+3)-3 a \sqrt{S}\right)}{\left(\pi  b^2+S\right) \left(b^2 \left(\pi  a \sqrt{S}-2 \pi ^{3/2}\right)+S \left(\sqrt{\pi } (8 P S+1)-2 a \sqrt{S}\right)\right)}.
\end{align}

\begin{figure*}[ht!]
\subfigure[]
{
\includegraphics[width=.3\textwidth]{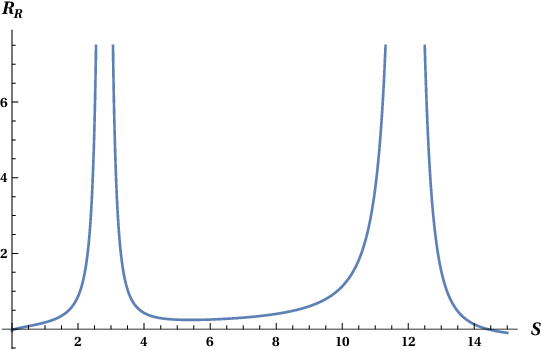}
\label{CRS1}
}
\subfigure[]
{
\includegraphics[width=0.3\textwidth]{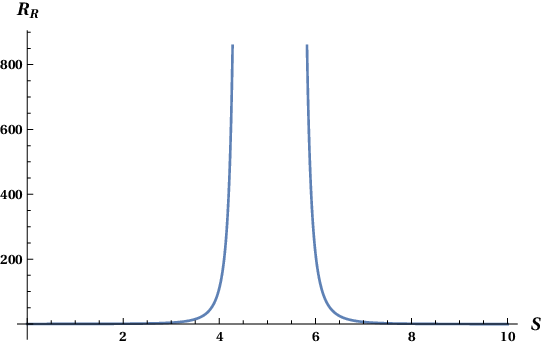}
\label{CRS2}
}
\subfigure[]
{
\includegraphics[width=0.3\textwidth]{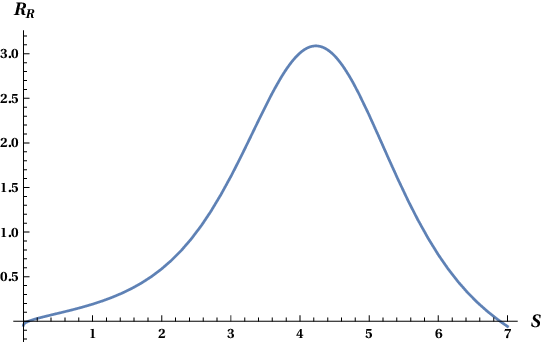}
\label{CRS3}
}
\\
\caption{Curvature divergence plots for Ruppeiner metric. In all three plots quintessence parameter and monopole charge are fixed ( $a=0.5$ and $\beta =1$). The value of pressure is $P=0.01$ in \ref{CRS1}, $P=0.01264$ in \ref{CRS2} and $P=0.0141$ in \ref{CRS3}.}\label{Ruppeiner RS}
\end{figure*}

The curvature tensor $R_R$ calculated from the above metric $g^R_{ij}$  is again a complicated expression like in the Weinhold case. The obtained curvature function is plotted against entropy $S$ to study the critical behavior (figure \ref{Ruppeiner RS} and \ref{CriticalR}). 

The figure \ref{CriticalR} shows that at the critical point $P_c=0.207$, there are multiple divergence around $S=0.06$ and $S=0.48$, which does not correspond to the critical  value of entropy $(S=0.32)$. From these multiple singularities for curvature scalar, it is difficult to identify the critical points from Ruppeiner geometry.  But, interestingly, Ruppeiner geometry indicates a phase transition even though it cannot identify the exact transition points (\ref{Ruppeiner RS}), like Weinhold geometry. This kind of anomaly was found in Kehagias-Sfetsos black hole \cite{GTDJanke2010}, Gauss-Bonnet Born-Infeld massive gravity theories \cite{Hendi2016, Hendi2015} and Einstein-Maxwell-dilaton black hole \citep{Wang2018}. In both Weinhold and Ruppeiner geometries, we note that the number of divergence points reduces of curvature scalar decreases when the pressure increases and gradually disappear.

\subsection{Geometrothermodynamics}
In this approach, the metric is constructed from a Legendre invariant thermodynamic potential and their derivatives with respect to the extensive variables. For geometrothermodynamics calculations, we will consider $2n+1$ dimensional thermodynamic phase space $\mathcal{T}$. 
This phase space is constructed using the coordinates $\{ \Phi , E^a, I^a \}$, where $\Phi $ is thermodynamic potential and $E^a$ and $I^a$ are extensive and intensive variables. Then Gibbs one-form is introduced as $\Theta = d \Phi -\delta _{ab} I^a E^b$, satisfying $\Theta \wedge (d\Theta) \neq 0 $. Defining a Legendre invariant metric $G$ on $\mathcal{T}$,
\begin{align}
G&=(d \Phi -\delta _{ab} I^a E^b)^2+(\delta _{ab} I^a E^b) (\eta _{cd} I^c E^d)\\
\eta _{cd}&=\textit{diag} (-1,1,......1).
\end{align}
$\mathcal{T}$, $\Theta$ and $G$ constitutes a Riemann contact manifold. Following this we define an $n$ dimensional Riemannian submanifold $\varepsilon \subset \mathcal{T}$, which is the space of equilibrium thermodynamic states (equilibrium manifold) via a smooth map $\varphi : \varepsilon \rightarrow \mathcal{T}$. The Quevedo metric, which is similar to Ruppeiner and Weinhold metric, is defined on this equilibrium submanifold using the inverse map $\varphi ^* (G)$.

\begin{equation}
g^Q = \varphi^*(G) =\left(E^c\frac{\partial \Phi}{\partial E^c}\right)
\left(\eta_{ab} \delta^{bc}\frac{\partial^2 \Phi}{\partial E^c \partial E^d}d E^a d E^d\right) 
\end{equation}

\begin{figure*}[ht!]
\subfigure[]
{
\includegraphics[width=.3\textwidth]{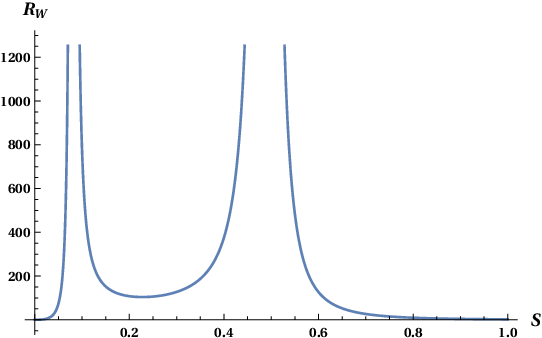}
\label{Criticalw}
}
\subfigure[]
{
\includegraphics[width=0.3\textwidth]{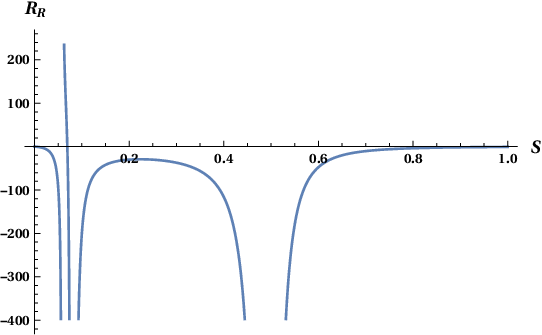}
\label{CriticalR}
}
\subfigure[]
{
\includegraphics[width=0.3\textwidth]{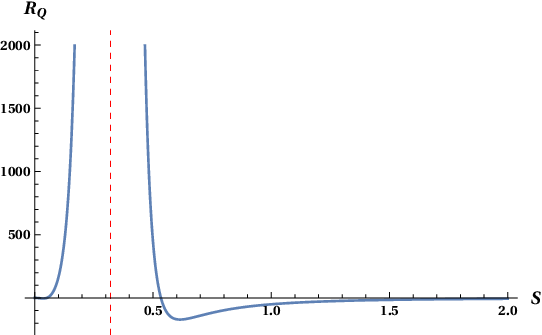}
\label{CriticalQ}
}
\\
\caption{Curvature divergence plots for Weinhold (figure \ref{Criticalw}), Ruppeiner (figure \ref{CriticalR}) and Quevado metric (figure  \ref{CriticalQ}) around critical point ($a=0.5,\beta =0.1$ and $P_c=0.207$).}\label{Critical}
\end{figure*}

In our case we consider a 5 dimensional phase space with the coordinates $Z_A =\{M, S, \beta, T, \Theta\}$, where $S$, $\beta$ are extensive variables and $T$, $\Theta$ are their dual intensive variables. Then we have the fundamental Gibbs one form as,

\begin{equation}
\Theta =dM-TdS-\Psi d \beta .
\end{equation}
Now we can write the Quevedo metric as follows,
\begin{equation}
g^{Q} = \left(S M_S + \beta M_\beta \right)\begin{bmatrix}
    -M_{SS}       & 0  \\
    0      & M_{\beta \beta} 
\end{bmatrix}.
\end{equation}
Using the Quevedo metric, we calculate the corresponding curvature, which is a complicated expression having the following form,

\begin{equation}
R_{Q}=\frac{f(S,\beta,P,a)}{g(S,\beta,P,a)},
\end{equation}
which is interesting as it has a diverging behaviour. Using the plots of curvature scalar $R_Q$, we investigate the divergence. In the figure \ref{CriticalQ}, we can see a divergence peaked at $S\approx 0.32$, same as in the specific heat case. Contrary to what we obtained in Weinhold and Ruppeiner geometries, the singular point of curvature scalar in geometrothermodynamics exactly matches the specific heat singular point. 

\section{Conclusion}
\label{sec:conclu}
We have studied the thermodynamics and thermodynamic geometry of a regular Bardeen-AdS black hole surrounded by quintessence. In the thermodynamic study, we observed a critical behavior from $P-v$ and $T-S$ plots. The critical values for pressure ($P_c$), volume $(v_c)$ and temperature ($T_c$) are obtained for $\omega=-1,-\frac{2}{3}$ and $-\frac{1}{3}$. The ratio  $\frac{P_c v_c}{T_c}$ decrease slightly  with  increase of $\omega$
from $-1$ to $-\frac{1}{3}$. It was found that the critical point depends on the quintessence parameters $a$ and $\omega$. More information about criticality was obtained from the specific heat plots. The discontinuity in the specific heat at $S=0.32$ indicates a phase transition of the system. For further confirmation, the Gibbs free energy is calculated and plotted against entropy. The presence of local minimum in $G-S$ plot manifests a thermodynamic instability. We analyzed the effect of quintessence in phase transitions by varying the state parameter $\omega$ in  $P-v$, $T-S$, $C_p-S$, and $G-S$ plots. 

Following the study of black hole phase transition in the thermodynamic approach, we carried out the geometrical investigation of the same. In the literature, it is a well-known fact that the divergence behavior of curvature scalar also reflects the existence of critical points. If we accept that the criticality of specific heat as the definition of phase transition, the thermodynamic geometry which shows divergence at the same point turns out to be the correct geometrical description of the same phenomena. For the metric under consideration, we found that, even though the Ruppeiner and Weinhold geometries reflect the singularity of curvature scalar, it can only be taken as the indication of phase transition but not the accurate description of the same, as the diverging points do not coincide with that of specific heat. There were multiple divergence and mismatch in the thermodynamic scalar of Weinhold and Ruppeiner geometries. This indicates an anomaly; to overcome this, we have used Quevedo's geometrothermodynamics. The main problem with Weinhold and Ruppeiner geometry is that they were not Legendre invariant and thus depend on the choice of thermodynamic potential. However, geometrothermodynamics being Legendre invariant, reproduces critical point exactly.

As future work, we would like to apply the thermodynamic approach formulated by S. Hendi \textit{et al.} \cite{Hendi2015}. The discrepancy among the different geometrical descriptions invalidating the critical behavior is still not clear conceptually; the solution to this may lie in the domain of quantum gravity. The underlying reason for the critical behaviour is the microscopic interaction in the black hole. The discrepancy between different geometric approaches can be resolved only with the proper understanding of the black hole microstructure. Even though there were successful attempts to count the microstates in supersymmetric black holes, the computation of the same for simple black holes like Schwarzschild is yet to be done. But still hopeful in research for a quantum gravity theory that can guide us in studying microstructure of black holes.


\acknowledgments
Authors A.R.C.L. and N.K.A. would like to thank U.G.C. Govt. of India for financial assistance under UGC-NET-SRF scheme.


  \bibliography{BibTex}

\end{document}